\documentclass[superscriptaddress, twocolumn, prl, aps]{revtex4-1}

\usepackage{graphicx}
\usepackage{amsmath, amsthm, amssymb, bm}
\usepackage{color}      

\begin{document}

\title{Comment on: ``Relating chain conformations to extensional stress in entangled polymer melts"}
\author{Wen-Sheng Xu}
\affiliation{Center for Nanophase Materials Sciences, Oak Ridge National Laboratory, Oak Ridge, Tennessee 37831, USA}
\author{Christopher N. Lam}
\affiliation{Center for Nanophase Materials Sciences, Oak Ridge National Laboratory, Oak Ridge, Tennessee 37831, USA}
\author{Jan-Michael Y. Carrillo}
\affiliation{Center for Nanophase Materials Sciences, Oak Ridge National Laboratory, Oak Ridge, Tennessee 37831, USA}
\affiliation{Computational Sciences and Engineering Division, Oak Ridge National Laboratory, Oak Ridge, Tennessee 37831, USA}
\author{Bobby G. Sumpter}
\affiliation{Center for Nanophase Materials Sciences, Oak Ridge National Laboratory, Oak Ridge, Tennessee 37831, USA}
\affiliation{Computational Sciences and Engineering Division, Oak Ridge National Laboratory, Oak Ridge, Tennessee 37831, USA}
\author{Yangyang Wang}
\email{wangy@ornl.gov}
\affiliation{Center for Nanophase Materials Sciences, Oak Ridge National Laboratory, Oak Ridge, Tennessee 37831, USA}


\maketitle

Based on non-equilibrium molecular dynamics simulations of entangled polymer melts, a recent Letter \cite{Robbins} claims that the rising extensional stress is \textit{quantitatively} consistent with the decreasing entropy of chains at the equilibrium entanglement length. We point out that exactly the opposite is true: the intrachain entropic stress arising from individual entanglement strands generally does not agree with the total ``macroscopic" stress.

We repeated the simulations of uniaxial extension for the $N=500$ and $k_{\mathrm{bend}}=1.5$ system, using the same approach employed in Ref. \cite{Robbins}, i.e., integrating the SLLOD equations of motion \cite{EvansMorriss} with the Generalized Kraynik-Reinelt boundary conditions \cite{NicholsonRutledge, Dobson}. The inverse Langevin function $L^{-1}$ \cite{Treloar}, originally derived for freely-jointed chains, can be used to estimate the entropic extensional stress in the large-strain limit:
\begin{equation}\label{eq:1}
\Sigma_e(n)=\frac{\rho k_B T}{C_n}\left\langle \frac{r(n)}{nb}L^{-1}\left[\frac{r(n)}{nb}\right]P_2(\cos \theta_n)\right\rangle,
\end{equation}
where $n$ is the number of bonds within a coarse-grained segment, $b$ is the bond length, and the characteristic ratio $C_n=\langle r^2(n)\rangle_0/nb^2$ is approximately equal to $C_{\infty}$ at $n=N_e$. Reference \cite{Robbins} used a similar equation. However, Eq. (\ref{eq:1}) in fact does not work for the semi-flexible chain model considered here: the maximum extension limit $nb$ can be exceeded in simulation for an \textit{individual} strand, particularly for small $n$ and large deformation, because of the ``soft" nature of the FENE bonds. In passing, we note that Ref. \cite{Robbins} did not distinguish the Legendre function $P_2(\cos \theta_n)$ and the nematic order parameter, i.e. $\langle P_2(\cos\theta_{n})\rangle$, in their Eq. (1), which is misleading.

\begin{figure}
\centering
\includegraphics[scale=0.5]{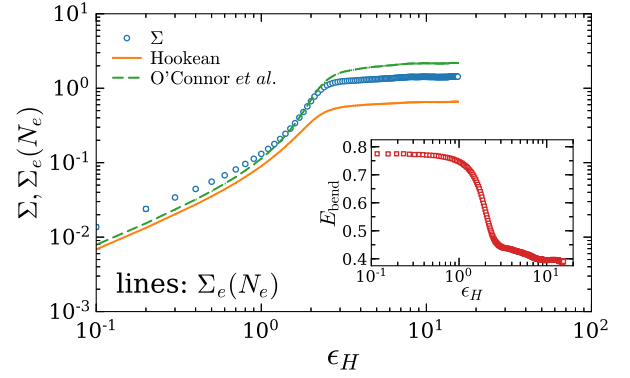}
\caption{Evolution of tensile stress $\Sigma$ and entropic tensile stress $\Sigma_e(N_e)$ as a function of Hencky strain $\epsilon_H$ for $N=500$ and $k_{\mathrm{bend}}=1.5$ at Rouse Weissenberg number $\mathrm{Wi}_R=25$. The (orange) solid line and (green) dash-dot line represent the entropic stresses $\Sigma_e(N_e)$ evaluated according to the Hookean spring law and Eq. (1) of Ref. \cite{Robbins}, respectively. The inset shows the bond bending potential during deformation.}
\label{fig:1}
\end{figure}

Figure \ref{fig:1} shows the evolution of tensile stress $\Sigma$ and entropic tensile stress $\Sigma_e(N_e)$ during a continuous extension simulation at $\mathrm{Wi}_R=25$, where $\Sigma_e(N_e)$ is evaluated according to the Hookean spring law and Eq. (1) of Ref. \cite{Robbins}. Regardless of the method, the entropic tensile stress at the entanglement length scale is substantially lower than the total stress $\Sigma$ at relatively small strains. This trend is true for all the other rates we examined, ranging from $\mathrm{Wi}_R=0.5$ to $50$. Unlike the case of large deformation, there should be no ambiguity in calculating the classical intrachain entropic stress \cite{DEbook} in the small-strain limit. The discrepancy between $\Sigma_e(N_e)$ and $\Sigma$ at relatively small strains clearly suggests that there is more to the story than the simple picture Ref. \cite{Robbins} paints. Generally speaking, ``quantitative" agreement between the entanglement strand entropic stress $\Sigma_e(N_e)$ and the total stress can be found only in a very limited range of $\epsilon_H$, even if $\Sigma_e(N_e)$ is computed through Eq. (1) of Ref. \cite{Robbins}. In fact, the steady-state data in Fig. 4 of Ref. \cite{Robbins} indicate a lack of ``quantitative" agreement at tensile stress higher than 0.1. Lastly, while the variation of the potential energy of the FENE bonds is indeed small even at high extension rates in these simulations, the bending energy does change substantially (inset of Fig. \ref{fig:1}) when the polymer coil is unraveled at large strains ($\epsilon_H \gg 1$). This is a direct violation of the assumption of purely entropic stress.

In summary, we show that the central result of Ref. \cite{Robbins} is premature: analysis of the full simulation trajectory reveals that the total extensional stress and the intrachain entropic stress at the equilibrium entanglement length generally does not agree with each other quantitatively, especially at relatively small deformation. Furthermore, in light of the ongoing debate about the origin of stress in entangled polymer melts in the recent literature \cite{Fixman, Gao3,  Ramirez2007, Likhtman2009, SussmanSchweizer}, the conclusion of Ref. \cite{Robbins}, which is based on an incomplete and questionable analysis of a limited range of the simulation trajectory, is particularly unconvincing.

\begin{acknowledgements}
This research used resources of the Oak Ridge Leadership Computing Facility at the Oak Ridge National Laboratory, which is supported by the Office of Science of the U.S. Department of Energy under Contract No. DE-AC05-00OR22725.
\end{acknowledgements}

\end{document}